# AOLI-- Adaptive Optics Lucky Imager: Diffraction Limited Imaging in the Visible on Large Ground-Based Telescopes

Craig Mackay*[a], Rafael Rebolo-López[b,f], Bruno Femenia Castellá[c], Jonathan Crass[a], David L. King[a], Lucas Labadie[d], Peter Aisher[a], Antonio Pérez Garrido[c], Marc Balcells[e], Anastasio Díaz-Sánchez[c], Jesús Jimenez Fuensalida[b], Roberto L. Lopez[b], Alejandro Oscoz[b], Jorge A. Pérez Prieto[b], Luis F. Rodríguez-Ramos[b], Isidro Villó[c],

[a]Institute of Astronomy, University of Cambridge, Madingley Road, Cambridge CB3 0HA, UK
[b]Instituto de Astrofisica de Canarias, C/ Via Lactea s/n, La Laguna, Tenerife E-38205, Spain and Departamento de Astrofísica, Universidad de La Laguna, La Laguna, Spain
[c]Universidad Politecnica de Cartagena, Campus Muralla del Mar, Cartagena, Murcia E-30202, Spain
[d]I. Physikalsiches Institut, Universität zu Köln, Zülpicher Strasse 77, 50937 Köln, Germany
[e]Isaac Newton Group, Apartado de Correos 321, Santa Cruz de la Palma, Canary Islands, Spain, E-38700
[f]Consejo Superior de Investigaciones Científicas, Spain

**ABSTRACT**

The highest resolution images ever taken in the visible were obtained by combining Lucky Imaging and low order adaptive optics. This paper describes a new instrument to be deployed on the WHT 4.2m and GTC 10.4 m telescopes on La Palma, with particular emphasis on the optical design and the expected system performance. A new design of low order wavefront sensor using photon counting CCD detectors and multi-plane curvature wavefront sensor will allow dramatically fainter reference stars to be used, allowing virtually full sky coverage with a natural guide star. This paper also describes a significant improvements in the efficiency of Lucky Imaging, important advances in wavefront reconstruction with curvature sensors and the results of simulations and sensitivity limits. With a 2 x 2 array of 1024 x 1024 photon counting EMCCDs, AOLI is likely to be the first of the new class of high sensitivity, near diffraction limited imaging systems giving higher resolution in the visible from the ground than hitherto been possible from space.

## 1. INTRODUCTION

The Hubble Space Telescope (HST) improved dramatically our ability to image the Universe, providing an 8-fold improvement in angular resolution over what can be delivered routinely by ground-based telescopes. There are many telescopes now that are substantially larger than the HST (2.4m), with a diffraction limit much smaller than the 0.12 arcsec images delivered by the HST. Unfortunately, the effects of atmospheric turbulence on image quality from ground-based telescopes have proved much harder to eliminate than had been hoped despite a major investment worldwide in adaptive optic (AO) technologies. There has been significant progress particularly in the near-infrared where the effects of turbulence are much less serious but it is still the case that in the visible no telescope on the ground has managed to deliver a resolution equal to that of the HST, even on a 2.4 m telescope solely by using AO.

However, HST angular resolution on HST size telescopes is routinely delivered in the visible by a technique called lucky imaging. This method was originally suggested by Hufnagel[1] in 1966 and given its name by Fried[2] in 1978. Images are recorded at high frequency to freeze the motion caused by atmospheric turbulence. A relatively bright reference star in the field allows image quality to be determined. The best fraction of images are shifted and added to give a combined image close to diffraction limited in their image quality when a fraction of ~5-30% is selected, the percentage depending on the atmospheric conditions at the time. An example of such an image is shown in Figure 1.

*cdm <at> ast.cam.ac.uk

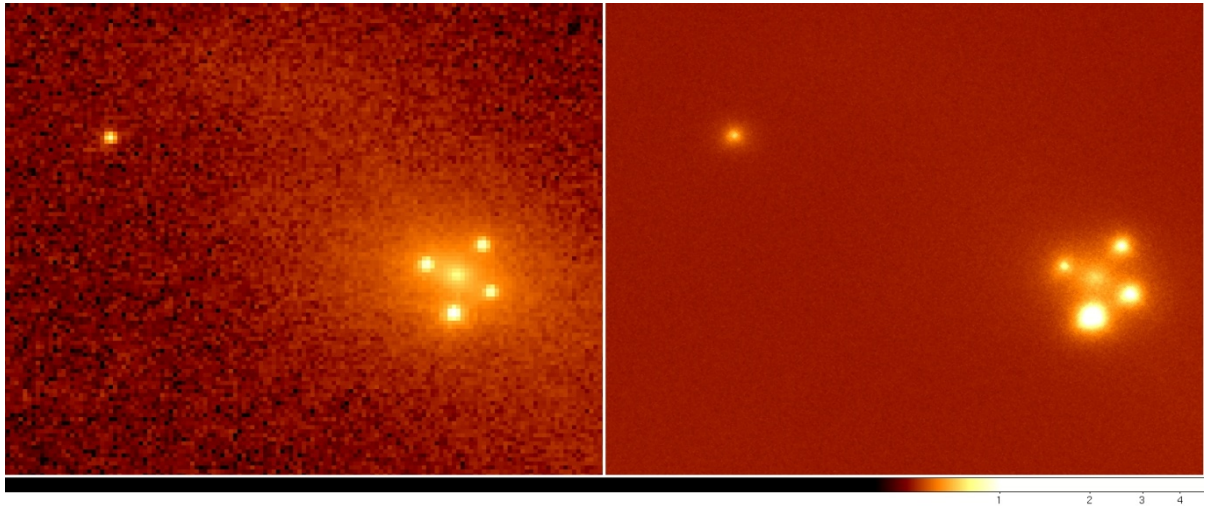

Figure 1: Images of QSO 2237+0305 (the Einstein Cross) gravitational lens. The light from a distant quasar is bent by a massive object in the core of a nearby Zwicky galaxy seen as the fuzzy object between four quasar images. The image on the left was taken with the HST Advanced Camera for Surveys (ACS) while the one on the right was taken by a Lucky Imaging camera on the 2.5m NOT telescope on La Palma. Microlensing within the lensing galaxy causes the relative brightness of the four images to change on relatively short timescales hence the difference in flux ratios.

The Lucky Imaging technique has become viable in recent years principally because of the development of electron multiplying CCDs (EMCCDs) particularly by E2V Technologies Ltd (Chelmsford, UK). These devices have most of the characteristics of conventional CCDs used routinely by astronomers and which are read out at low speed. However, a modification of the device output register[3] provides internal amplification by large factors, up to a few thousand times. At high pixel rates of around 30 MHz and readout noise of around 100 electrons RMS this gain allows individual photons to be detected with good signal-to-noise. This permits imaging at high frame rates without any read noise penalty. In photon counting mode it is possible to operate with almost all of the quantum efficiency of a conventional CCD making these devices particularly attractive for a range of high-speed imaging and spectroscopy applications in astronomy and other research areas.

Theoretically, large telescopes can deliver sharper images than HST. However, the probability of the Lucky Imaging technique delivering near diffraction limited images becomes vanishingly small for telescopes significantly larger than the HST[2]. This is because the number of turbulent cells across the diameter of the telescope is too large for there to be a significant chance of a relatively flat wavefront (and hence a near diffraction-limited image) across the aperture of the telescope. By increasing the diameter of the telescope we bring in the effects of yet larger scales of atmospheric turbulence. In principle, if we could eliminate the largest turbulent scales where most of the power in the atmospheric turbulence resides[4] then the probability of recording a sharp image will increase. Essentially, eliminating one turbulent scale reduces the phase variance across that scale so that the characteristic cell size, $r_0$ (defined as the scale size over which the variance is ~1 radian$^2$) is increased. Provided enough of the large turbulent scales are removed, the corrected $r_0$ will be large enough so the number of cells across the diameter of the telescope is similar to those typically encountered with an uncorrected 2.4 m aperture.

To test this, one of the Cambridge Lucky Imaging cameras was mounted on the Palomar 5-m telescope behind the low order adaptive optics system PALMAO[5]. The results were very exciting in that they produced the highest resolution picture ever taken of faint targets in the visible from any telescope on the ground or in space (Figure 2). Images with an angular resolution of 35 milliarcseconds in I band (770 nm) were obtained, a resolution more than 3 times that of the HST (which is undersampled in the visible and so only produces images of about 0.12 arcsec resolution). The performance of this combination of Lucky Imaging plus low order adaptive optics is compared with that of the Advanced Camera for Surveys on HST in Figure 3. Other groups in Europe and in the US have contributed to the scientific development of optical Lucky Imaging, in view of multiplicity studies of sub-stellar objects, low- and high-mass stars[6,7,8,9]. Instruments such as FastCam and AstraLux have worked on 2-m and 4-m class telescopes and used the full diffraction-limited resolution in the z'- and I- bands for the studies mentioned above. In the US, projects aiming at developing visible AO systems are also on-going[10].

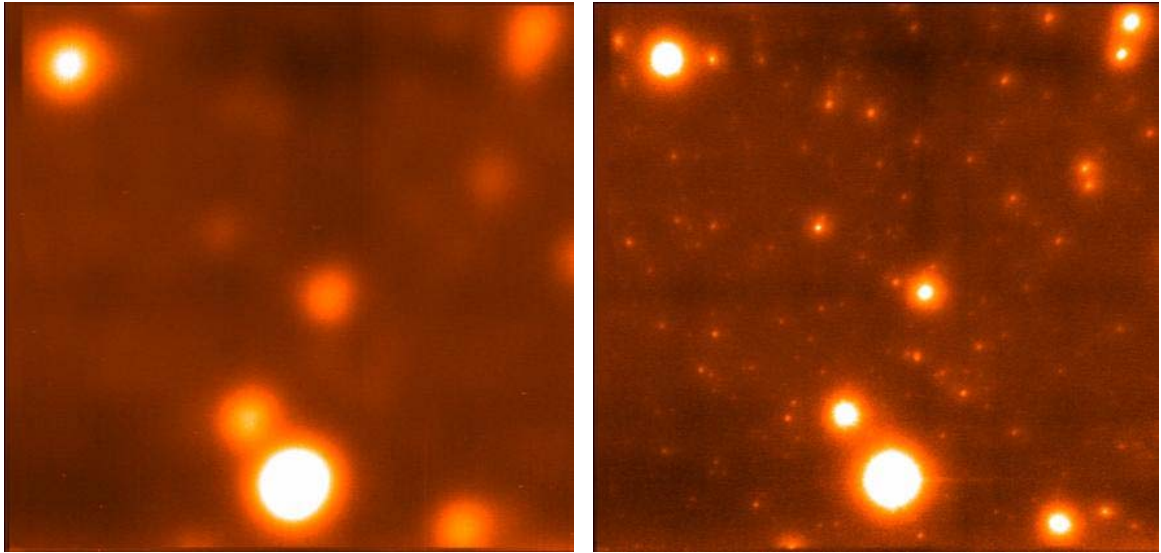

Figure 2: The core of the globular cluster M 13 in I band observed on the Palomar 5-m telescope. The left image is with natural (~0.65 arcsec) seeing, and the right with the Lucky Imaging Camera behind the low order PALMAO adaptive optic system on the Palomar 5-m telescope. The resolution in this image is about 35 milliarcseconds or approximately 3 times that of the (undersampled) Hubble Space Telescope. The total field of view is about 10 x 10 arcseconds. This is the highest resolution image of faint objects ever taken in the visible or infrared anywhere from space or from the ground. The isoplanatic patch size is clearly large, much greater than 10 arcsec.

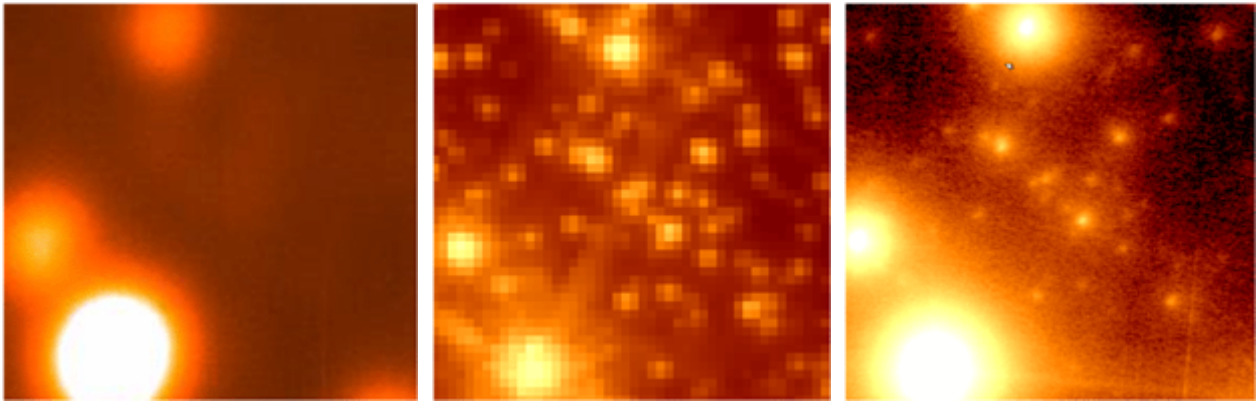

Figure 3: Comparison images of the core of the globular cluster M 13. On the left the image with natural (~0.65 arcsec) seeing on the Palomar 5-m telescope, the Hubble Advanced Camera for Surveys with ~120 milliarcsecond resolution (middle) and the Lucky Camera plus Low-Order AO image with 35 milliarcsecond resolution (right).

Although these results were dramatic, PALMAO, in common with many current adaptive optics systems, requires a very bright reference star for the Shack-Hartmann wavefront sensor. Shack-Hartmann AO systems generally need reference stars of I~12-14 magnitude, and these are very scarce[11]. High order AO systems also have a very small isoplanatic patch in the visible of only a few arcseconds in diameter[12]. This means that the technique can only be used over a very small fraction of the sky (much less than 1%). Laser guide stars are being tried increasingly but they also have problems in delivering images with ~0.1 arcseconds accuracy. The net effect is that it is very hard to achieve a resolution better than 0.1 arcseconds even in the near infrared on large telescopes although this has been achieved in a limited number of instances.

Studies of the sensitivity of wavefront sensors by Racine[13] suggest that curvature sensors deployed on telescopes are significantly more sensitive than Shack-Hartmann sensors particularly when used for relatively low order turbulent correction. Olivier Guyon[14] simulated the performance of pupil plane curvature sensors and showed that they are very attractive in general and, when combined with EMCCDs ought to give a substantial improvement in sensitivity. Such an improvement would allow the use of a much larger fraction of the sky, and the new AOLI (Adaptive Optics Lucky Imager) instrument is designed to combine a low order photon counting curvature sensor with a high sensitivity photon counting science camera working in Lucky Imaging mode. AOLI is a collaboration between the Instituto de Astrofisica de Canarias/Universidad de La Laguna (Tenerife, Spain), the Universidad Politecnica de Cartagena (Spain), Universität zu Köln (Germany), the Isaac Newton Group of Telescopes (La Palma, Spain) and the Institute of Astronomy in the University of Cambridge (UK). This paper will describe the overall design of the instrument, its optical configuration and predicted performance, and also describe improvement and developments in a range of areas key to the success of the instrument.

## 2. AOLI: GENERAL CONFIGURATION

The AOLI instrument consists of a non-linear curvature wavefront sensor and a low order adaptive optics wavefront corrector using a deformable mirror in conjunction with a wide-field, array detector Lucky Imaging camera. It is designed specifically for use on the WHT 4.2-m and the GTC 10.4-m telescopes on La Palma but it could be used on almost any large telescope without major modification.

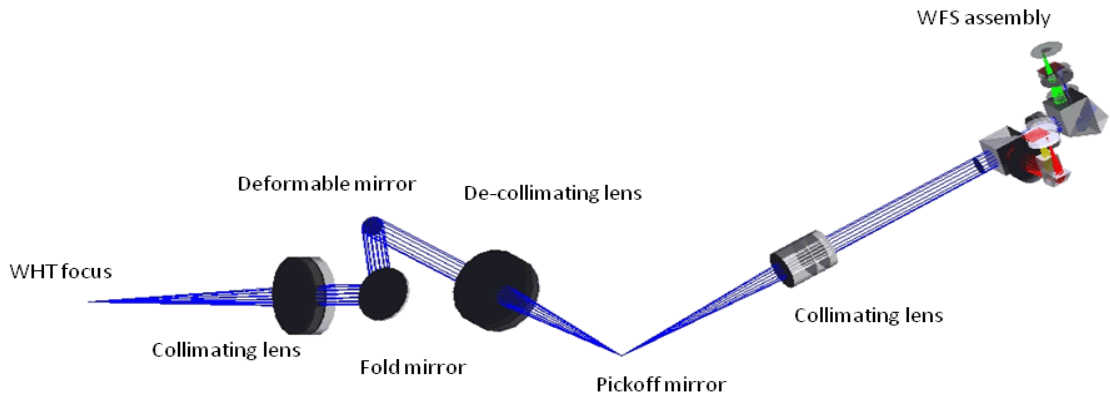

Figure 4: A solid model of the optical arrangement of the curvature sensor. The common optics take light from the WHT focus to the pickoff mirror where the light from the reference star is diverted to the WFS assembly, and is split into 2 beams. One beam is deflected by a cube beamsplitter and again split to give a pair of near-pupil planes on to one detector. The other beam continues through the cube beamsplitter and providing more widely separated pupil planes onto a second detector. The wavefront errors are determined by photon counting electron multiplying CCD cameras developed in Cambridge running at about 100 Hz frame rate. The data from these detectors are processed to derive the wavefront errors and these signals are then used to drive the deformable mirror to remove the largest scales of turbulence from the wavefront. The science beam continues straight on at the pickoff mirror plane to the science imaging detector (see Figure 6) which is then used in conventional Lucky Imaging mode.

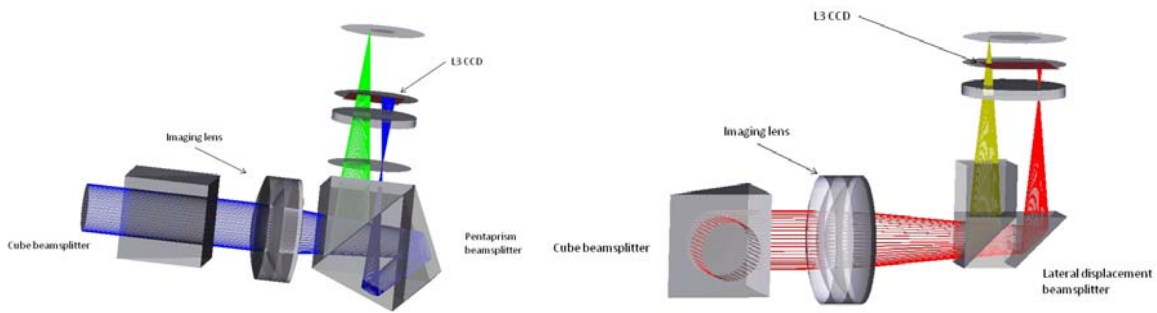

Figure 5: A more detailed view of the wavefront sensor assembly, showing the form of the two beamsplitters which deliver the four beams to the two EMCCds.

AOLI includes an atmospheric dispersion corrector, essential for work away from the zenith, forward of the focal plane in Figure 4 (the ADC is not shown in the figure). In Figure 4 the science beam goes from the telescope straight down to the science detector while a pickoff mirror deflects light from the reference star to a pair of beam splitters and photon counting detectors. The reference star is located on the optical axis of the telescope and a pickoff mirror mechanism allows reflecting spots of different sizes and optical densities (for different observing conditions and different targets) to divert light to the wavefront sensor. The light from the telescope is reflected via a deformable mirror set in a pupil plane of the telescope which allows the curvature errors determined by the curvature sensor sub-system to be corrected directly. The deformable mirror is manufactured by ALPAO (Gières, France) with 97 elements giving 11 deflectors across the pupil. It will allow correction of wavefront errors on scales of > ~1.0m on the 4.2 m diameter WHT telescope. Our simulations[15] suggest that this will then give us a Lucky Imaging selection percentage under typical/good conditions of about 25-30% in I band. Additional simulations suggest that we will need to use the higher resolution (19 x 19 array) deflector system from ALPAO in order to achieve good enough correction on the GTC 10.4 m telescope. Our approach is to develop a system optimised for the WHT that may be modified and re-deployed quickly in order to demonstrate the technologies as convincingly as possible. Its subsequent deployment to the GTC will then follow.

The science camera is a simple magnifier using custom optics to give diffraction limited performance. The camera is optimised for the 500nm to 1 micron wavelength range. The diffraction limit of the WHT (GTC) at 0.8µm (I band) is about 40 (15) mas and the camera offers a range of pixel scales of between 6 and 60 mas. The camera uses an array of 4 photon counting, electron multiplying, back illuminated CCD201s manufactured by E2V Technologies Ltd, each 1024 x 1024 pixels. As the CCDs are non-buttable we use an arrangement similar to that of the original HST WF/PC (see Figure 5). Four small contiguous mirrors in the focal plane are slightly tilted and then individually reimaged on to a separate CCD. Each CCD has its own filter wheel. This allows the use of a narrowband filter, for example, for the science object with a broad filter for the reference star. The configuration allows a contiguous region of 2000 x 2000 pixels giving a field of view of from 120 x 120 arcsec down to 12 x 12 arcseconds depending on the magnification selected.

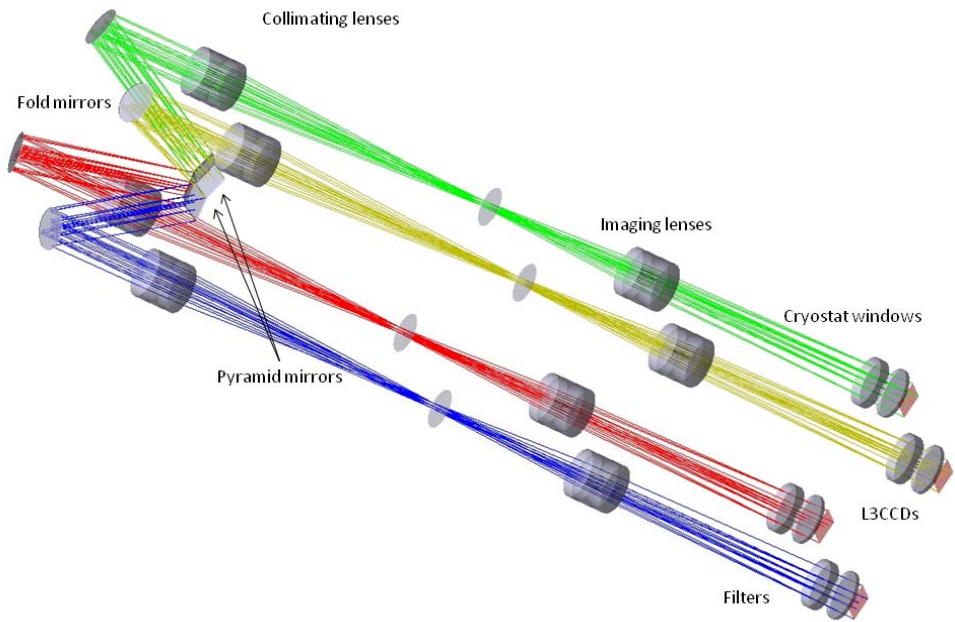

Figure 6: The science camera optical arrangement whereby the light from a single area of sky is split on to four separated and non-buttable CCDs. The magnified image (optics not shown) of the sky is projected onto a pyramid of four mirrors that reflects the light on to relay mirrors and via reimaging optics onto four electron multiplying detectors. This structure was suggested by the design of the original widefield/planetary camera installed on the HST.

The CCDs are back illuminated (thinned) with very high quantum efficiency (peak >95%) from E2V Technologies. Custom electronics developed in Cambridge give up to 30 MHz pixel rate, and 25 frames per sec. Higher frame rates are possible. A readout format of 2000 x 100 pixels gives ~200 fps, allowing high time resolution astronomy as well with the instrument. The data (~220 MBytes/sec continually) are streamed via the host computer to high-capacity disk drive systems after lossless compression. The host computer performs basic lucky imaging selection, allowing image quality to be assessed while the exposure is progressing. The construction of the unit will be kept relatively low-cost. The instrument will be mounted at Nasmyth focus on an optical bench behind the WHT image rotator. On the GTC it is probable that the instrument would be mounted on one of the folded Cassegrain ports.

### 3. NON-LINEAR PHOTON COUNTING CURVATURE WAVEFRONT SENSORS

Most adaptive optics development programmes today are targeted at achieving very high degrees of correction of the incoming wavefront errors. Most use Shack-Hartmann sensors and therefore require very bright and scarce reference stars (typically I~12-14 magnitude). Low order correction with much fainter reference stars may be achieved more easily with a curvature sensor. These work by taking images on either side of a conjugate pupil plane and looking at changes in the intensity of illumination as the wavefront passes through the pupil. A part of the wavefront that becomes fainter as it goes through the pupil must correspond to a part of the wavefront that is diverging while if it becomes brighter it is converging. Racine[13] has shown that curvature sensors actually deployed on telescopes are typically 10 times (2.5 magnitudes) more sensitive than Shack-Hartmann sensors for the same degree of correction. In addition they are very much more sensitive again when used for low order correction as the system cell size is dynamically increased[8] and the wavefront sensor readout rate/integration time may be significantly reduced. Correction with coarse cell sizes allows averaging sensor signals over significant areas of the curvature sensor.

We have been working to simulate what could be achieved with a low order non-linear curvature sensor on a large (5-10 m) ground-based telescope. On the WHT the angular resolution will be very similar to that obtained on our Palomar run described above. On the GTC it will be another factor of 2.5 better. This will give us resolutions of typically 15-40 milliarcseconds in the visible to I-band range. Our simulations suggest that we will need a reference star of about 17.5-18.0 on the WHT and about 18.5-19.0 mag on the GTC. This will allow us to find reference stars over nearly all the sky even at high galactic latitudes (>85%)[5]. Our simulations indicate that partial compensation will also be possible with yet fainter stars.

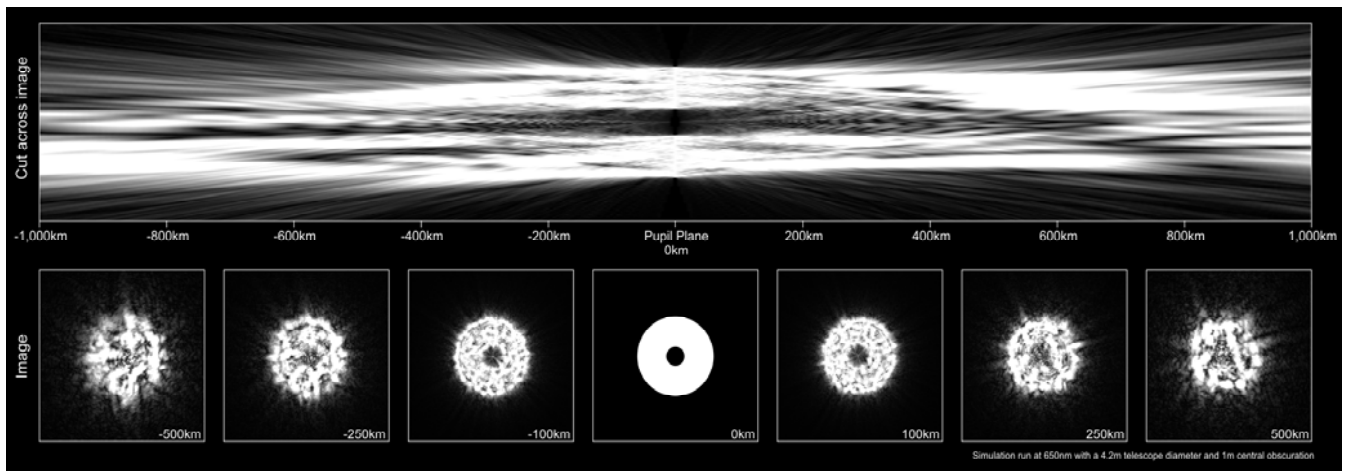

Figure 7: The propagation of light through the pupil of the telescope. At the pupil the illumination is uniform. On either side of the pupil the intensity breaks into speckles whose size is ~ the diffraction limit of the telescope. Further out the structure is dominated by the lower order turbulent components which we wish to remove from the wavefront. The wavefront propagation is substantially achromatic in low order allowing the use of broad response bands with a curvature sensor[17]. We make no attempt to remove the high order components, relying on the lucky imaging technique to provide that selection. We use four out-of-pupil imaging planes with photon counting high-speed CCD cameras already developed in Cambridge.

The propagation of the wavefront through the pupil is a non-linear process so that the best sensitivity comes by using four pupil planes[16,17], one pair close to the pupil on either side and the second pair further from the pupil, also on either side. Figure 7 shows a cross-section through a typical turbulent wavefront as it goes through the pupil. A critical aspect of our design of curvature sensor is its ability to work at very low light levels and to be stable at the lowest light levels we wish to use. This requires the non-linear wavefront propagation to be computed using Bayesian statistical methods. We have already developed high-speed wavefront reconstruction algorithms needed to run very fast in order to avoid unwanted latencies in the wavefront correction loop.

## 4. ENHANCED EFFICIENCY LUCKY IMAGING

Classical lucky image selection[18] starts with a large number of frames containing a reference star of "adequate" signal-to-noise, ideally more than about 200 detected photons per frame (corresponding to I ~ 16m on a 2.5-m telescope). The position of the brightest speckle of the reference star is used to establish the shift to be made to each frame to move it into registration and then add it into the total. By selecting a subset of the frames which have the sharpest images indicated by the brightness of the brightest speckle[18], the angular resolution may be improved significantly. Even with 100% selection the shift and add procedure essentially eliminates the contribution to image smear caused by tip-tilt due to turbulence or telescope pointing errors, and will typically double the resolution of the telescope without any image selection. The signal-to-noise of the reference star is important since at low signal levels the brightness of the central speckle can depend as much on the statistics of photon arrival times as it can on the intrinsic variance of the incoming wavefront. For critically sampled or undersampled images it is generally necessary to use subsampling so that each frame may be located with an accuracy of a fraction of a pixel so as not to compromise the accumulated angular resolution.

The highest resolution images are obtained from choosing the smallest percentage of frames which have the sharpest point spread functions (PSFs). In many cases it is clear that slightly poorer images are smeared in one direction and yet still have the full resolution in the orthogonal direction. Garrel, Guyon & Baudoz[19] (GGB) proposed that the lucky image selection might be done better in Fourier space than in image space. They describe their method and a series of simulations representing the predicted performance of their method on the 8 m Subaru telescope with 0.6 arcseconds seeing. This method enables the full resolution information to be used to improve the signal-to-noise while preserving overall angular resolution when more frames are combined, overcoming one of the less satisfactory aspects of the Lucky Imaging method, which only delivers high-resolution when a large fraction of the images are discarded. We have developed this proposal further and found that it works well both with conventional Lucky Imaging datasets and Lucky Imaging plus low order AO datasets.

The two-dimensional Fourier transform of an astronomical image which has been shifted so that the zero frequency component is in the middle of the transform array shows a central peak that is broader in any direction that corresponds to better angular resolution. Classic Lucky Imaging relies on attributing a quality rating to whole images on the basis of the sharpness of the reference star in each frame. In the Fourier plane, GGB suggest ordering each element in the complex (u,v) plane in order of amplitude. The percentage selection is then made amongst the corresponding elements in each frame of the complete sequence. For any particular element in the complex (u,v) plane, the highest amplitude (u,v) elements will be derived from different frames and so each (u,v) location needs to be ordered by amplitude independently. For example, the 10% selection frame consists of the average of the highest amplitude 10% of (u,v) elements recorded for each and every (u,v) location independently. This 10% Fourier transform is then inverse transformed to give the 10% selected output image. That image may be compared directly with the 10% classic Lucky Imaging selected image, a detailed account of these methods is given elsewhere[20].

We have used a variety of datasets both with conventional Lucky Imaging and adaptive optics assisted Lucky Imaging. The results are consistent with the GGB simulations. We find that the key quality indicators, specifically photometric and astrometric accuracy of any image are preserved. Examples of the data produced are given in Figures 8 and 9.

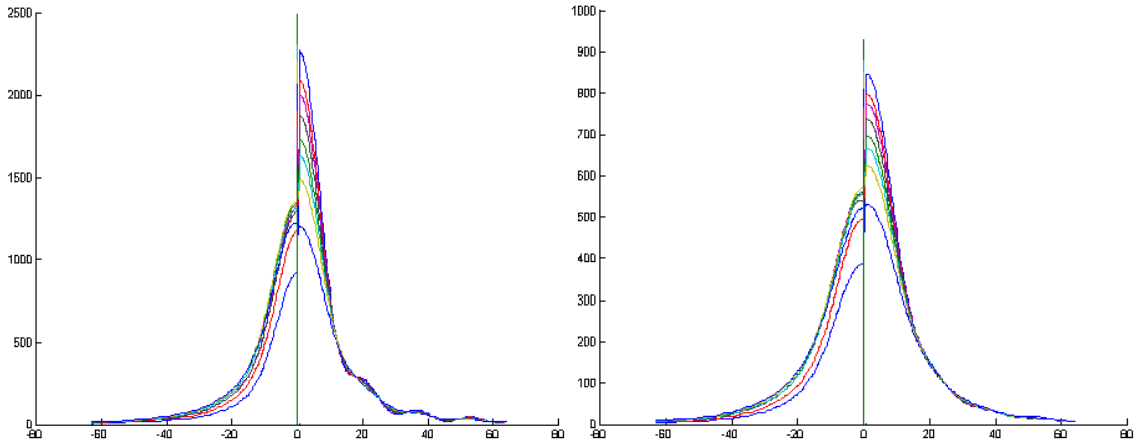

Figure 8: One-dimensional cuts through 2 separate stars. The profiles correspond from the top to the bottom curve in each half with Lucky Imaging selections of 1%, 3%, 5%, 10%, 20%, 30%, 50%, and 100%. The left-hand half of each plot shows the profile of the star as synthesised using conventional Lucky Imaging where images are simply selected on the basis of the sharpness of the reference star. The right-hand half of each plot shows the profile obtained with the Fourier method. It is clear that the synthesised star is sharper and has a significantly higher Strehl ratio. The left-hand star is very close to the reference star while the right-hand star is at an angular distance of about 25 arcseconds. It is clear that the method does not compromise the size of the isoplanatic patch.

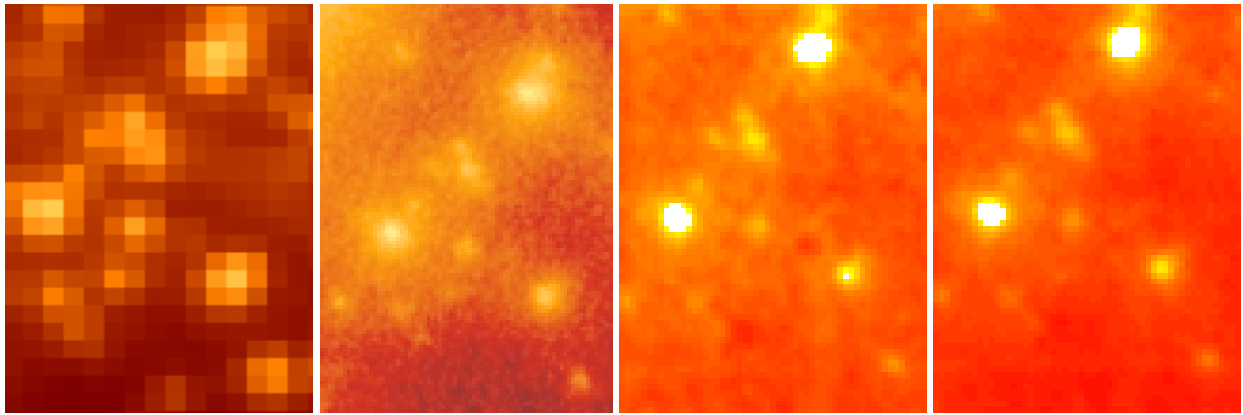

Figure 9: The above images demonstrate the effectiveness of the Fourier synthesis method on real high-resolution data (the same dataset used to produce Figure 2 and 3). The left-hand box is the Hubble Space Telescope ACS camera image, the next one is the lucky image synthesised in the conventional way with 10% selection (part of Figure 3). The right hand two images are formed with the Fourier synthesis Lucky Imaging method at 20% and 50% selection. The higher resolution of these images is clearly visible. The background halo around each star image is significantly suppressed in these images which cover approximately 2.0 x 1.5 arcseconds.

## 5. WAVEFRONT RECONSTRUCTION ALGORITHMS AND PROCEDURES

Shack-Hartmann AO sensors use a lenslet array to sample the wavefront over a rectangular grid of positions in the pupil plane. A reference star is imaged and the deflections of the sub images relative to the regular grid defined by the lenslets allows the local wavefront gradient to be derived. As the number of lenslets used in high order Shack-Hartmann systems increases, the computational complexity of inverting the measured positions of the sub images and turning this into the appropriate drive signals for the wavefront corrector to correct the wavefront becomes rather great. These calculations need to be completed and the wavefront correction applied before the phase errors in the incoming wavefront have changed.

With a non-linear curvature sensor, computing the wavefront curvature from the four near-pupil images is potentially difficult and time consuming. The optical arrangement (Figure 4) shows that pairs of near-pupil images are projected onto a single EMCCD. The second pair of near-pupil images is recorded with a second EMCCD. The images read out are typically 256 x 256 pixels and these are read out at approximately 100 Hz frame rate.

The four images recorded are, of course, intensity images whereas we really want to determine the phase (the curvature) of the incoming wavefront. Wavefront reconstruction with curvature sensors has been demonstrated by Mateen et al.[21] who used the Gerchberg-Saxton algorithm[22] and which provides convergence but it is very much slower than other procedures. Indeed they find the Gerchberg-Saxton algorithm to be a 10-100 slower than would be necessary for on-sky work. Feinup[23] has shown that input-output algorithms will generally converge much faster than error-reduction algorithms such as the Gerchberg-Saxton algorithm. More recently, Aisher et al.[24] have experimented with modifications of some of the Feinup algorithms to the case with four planes.

Critical for AOLI is the reliability of achieving fast convergence, particularly at the lowest signal levels which is where the scientific rewards of working with AOLI will be greatest. An example of the performance predicted is shown in Figure 10. It is worth remarking that the low order AO corrector needs to work on much slower timescales than high order ones. The wind crossing time of the ~1-1.5 metre cell size when the median windspeed is ~8 km/sec is ~150 ms. Although we will record wavefront data at ~100 Hz and be able to carry out the reconstruction comfortably in a few milliseconds the effective control loop will be running at typically 10-20 Hz. At the lowest light levels we will only be able to reconstruct the wavefront by averaging across the near-pupil images spatially and temporally to maximise sensitivity. The speed of the reconstruction algorithms developed, together with the use of 4 near pupil planes is high enough to allow several parallel processing strategies to be run simultaneously and allowing the best one to be used. We note in passing that the advantages that might be imagined to be gained by using graphics processor unit cards appear to be slight as the time needed to transfer data in and out of those cards from the computer host memory greatly reduces the overall efficiency in this particular application.

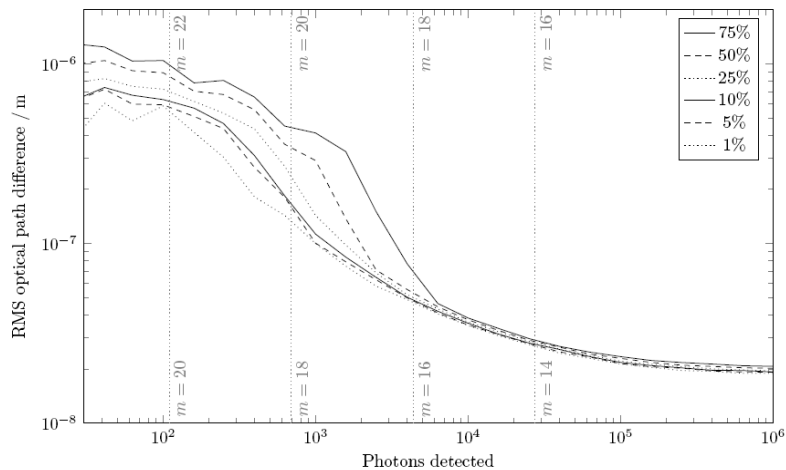

Figure 10: Simulated low light performance of the AOLI curvature wavefront sensor using an input-output algorithm. Vertical lines show the I-band magnitude required for a given photon count with the CWFS running in photon counting mode at 10 Hz for D = 4.2m (lower label), and D = 10.4m (upper label). The different curves show the RMS optical path differences achieved using different Lucky Imaging selection percentages.

One of the advantages of this approach to wavefront sensing is that we will know in very great detail what the quality of the image wavefront was at every moment that we record of scientific image. Those data enables a more precise assessment of image quality than is possible even in theory simply by looking at the point spread function in a recorded image. So we envisage that the lucky image selection may well be done on the basis of these wavefront reconstructions rather than by using the conventional Lucky Imaging strategies. We further will have the opportunity to examine the application of deconvolution techniques since we will know very accurately the true wavefront structure that gave rise to each of our science images.

## 6. CONCLUSIONS

In many ways the holy grail of ground-based observing in the visible is to develop techniques that enable us to observe close to the diffraction limit of large telescopes over the full sky with good observing efficiency. With AOLI we believe that we will be able to make substantial progress in that direction. The availability of high-speed, high efficiency, photon counting detectors has transformed our capacity to build instruments capable of working with the rapidly changing atmosphere and using very faint reference stars to let us optimise our recording of light from the sky. The combination of these EMCCDs with low order curvature wavefront sensors (also using EMCCDs) will allow a new generation of astronomers to explore the Universe with as big a step change in resolution as Hubble provided over 20 years ago. Hubble provided an eight-fold improvement over the typical ground-based image resolution of ~1 arcsec to give images of ~0.12 arcsec resolution. We have already demonstrated a further improvement over Hubble with the Palomar 5 m telescope by imaging with ~0.035 arcsec resolution. AOLI in the visible on the GTC 10.4m telescope has the diffraction limit of eight times better than HST of ~0.015 arcsec resolution, roughly 60 times the resolution when limited by atmospheric turbulence. There is every expectation that by making such high resolution images and spectra available more routinely many fields of astronomy will be revolutionised yet again.